\documentclass[11pt,twoside]{article}
\usepackage{asp2004}
\usepackage{epsf}
\usepackage{psfig}
\begin{document}

\title{The solar chemical composition}

\author{Martin Asplund$^1$}
\affil{$^1$Research School of Astronomy and Astrophysics, 
Australian National University,
Cotter Road, Weston 2611, Australia}

\author{Nicolas Grevesse$^2$}
\affil{$^2$Centre Spatial de Li\`ege, Universit\'e de Li\`ege,
avenue Pr\'e Aily, B-4031 Angleur-Li\`ege, Belgium 
and Institut d'Astrophysique
et de G\'eophysique, Universit\'e de Li\`ege, All\'ee du 6 Ao\^ut, 17, B5C, 
B-4000 Li\`ege, Belgium}

\author{A. Jacques Sauval$^3$}
\affil{$^3$Observatoire Royal de Belgique, avenue circulaire, 3,
B-1180 Bruxelles, Belgium}

\begin{abstract}
We review our current knowledge of the solar chemical composition as determined
from photospheric absorption lines. In particular we describe the recent
significant revisions of the solar abundances as a result of the application
of a time-dependent, 3D hydrodynamical model of the solar atmosphere
instead of 1D hydrostatic models.
This has decreased the metal content in the solar convection zone by almost
a factor of two compared with the widely used compilation by Anders \& Grevesse (1989). 
While resolving a number of long-standings problems,
the new 3D-based element abundances also pose serious challenges, most notably
for helioseismology.
\end{abstract}

\section{Introduction}

The chemical composition of the Sun is one of the most important
yardsticks in astronomy with implications for basically all fields
from planetary science to the high-redshift Universe.
Standard practise is to compare the element content in a cosmic
object with the corresponding value for the Sun using the
normal logarithmic abundance scale
$[{\rm X/H}] \equiv  {\log} \left(N_{\rm X}/N_{\rm H}\right)_* -
                     {\log} \left(N_{\rm X}/N_{\rm H}\right)_\odot \equiv
                     {\log} \epsilon_{\rm X, *} - {\log} \epsilon_{\rm X, \odot}$.
Abundance ratios such as [C/Fe] are defined correspondingly.
Hydrogen is the chosen reference element since it is the most
common element in the Universe as a whole and that it normally
dominates the continuous opacity in the formation of the stellar spectra.
The solar abundances are, however, not only of interest as a reference point
in astronomy but are also of profound importance for the understanding of
our own solar system and the solar interior.

Most of our knowledge of the overall abundance distribution of the elements in
the solar system originates in two
quite different sources: the solar photospheric spectrum and pristine
meteorites, in particular the so-called C1 chondrites\footnote{Important and 
complementary information is provided by
studies of for example the solar corona, solar wind, solar energetic particles,
solar flares and lunar soil, which
however will not be discussed here.}
 (e.g. Anders \& Grevesse 1989).
Each has their own advantages and disadvantages. The abundances of a
very large number of elements can be determined using absorption lines
in the solar spectrum. However, since the results are dependent on having realistic
models of the solar atmosphere and the line formation process the accuracy
can only be expected to be good at the 10\% (0.04\,dex) level at best.
Isotopic abundance information is only available in very rare instances.
Model-independent meteoritic isotopic abundances can be measured to exquisite accuracy
in the laboratory but only a very small subsample of meteorites do not show
effects of fractionation. In addition, elements like H, He, C, N, O and Ne 
are all volatile and hence
depleted in the meteorites. It is therefore clear
that the two methods complement each other well.
The present review will describe our knowledge of the solar photospheric
chemical composition, in particular in light of recent significant revisions of
these abundances following the application of the new generation of 
three-dimensional (3D) hydrodynamical
model atmospheres to the problem.

\section{Model Atmospheres, Line Formation and Other Obstacles}

While the chemical composition of a star is encoded in its spectrum, it is not
immediately extractable without first modelling the stellar atmosphere and the
line formation process. Since stellar element abundances are thus not observed
but derived they are by necessity model-dependent.
As more realistic modelling become available the estimated 
stellar abundances may therefore change with time. Equally important is
improved atomic and molecular data, which is nicely exemplified by 
Sneden et al. in these proceedings. 
Obviously progress in both of these areas is vital.
Indeed they are also complementary as
more realistic modelling can uncover shortcomings in the input physics and vice versa.

Traditionally, standard
stellar abundance analyses employ time-independent, one-dimensional (1D)
model atmospheres in hydrostatic equilibrium. Such models basically come in
two types: theoretical and semi-empirical. In the former the temperature structure
is computed assuming flux constancy, i.e. the same energy is transported through
each atmospheric layer. The energy is assumed to be carried solely by radiation
or convection. For simplicity, the convective energy flux is approximated with
the local mixing length theory (B\"ohm-Vitense 1958) or some close relative thereof
(e.g. Canuto \& Mazzitelli 1991). In the absence of convection, radiative equilibrium
holds. It is crucial to include the effects of blanketing from spectral lines
when integrating the radiative transfer equation over all wavelengths
in order to obtain a realistic temperature structure. LTE is normally assumed
for the computation of the atomic populations and the radiative transfer.
Some of the most widely-used theoretical 1D model atmospheres for late-type stars are those by
Kurucz (1979), the {\sc marcs} consortium (Gustafsson et al. 1975) and the Phoenix
team (Hauschildt et al. 1999), which are all continuously improving their models,
mainly through the inclusion of better and more complete atomic and molecular physics.

In a semi-empirical model
atmosphere the temperature is deduced from observations
through an inversion procedure, normally using spectral lines of different 
line formation depths and/or center-to-limb variations.
It therefore follows that the atmosphere in such a model
does not need to obey flux constancy nor is it necessary to estimate the radiative
and convective energy transport.
The danger with such a method is that the line formation calculations may be faulty,
for example due to the assumption of LTE or the use of
erroneous line data. This would directly
translate to an incorrect temperature structure. Hydrostatic equilibrium
is assumed also in semi-empirical model atmospheres.
Likewise, a knowledge of the equation-of-state is necessary to relate
temperature, gas pressure and electron pressure.
Semi-empirical atmosphere modelling has only been tried relatively few times for
stars other than the Sun (e.g. Allende Prieto et al. 2000) 
but is a relatively common practise in
solar analyses. Two of the most widely used solar semi-empirical models are the
VAL-3C (Vernazza et al. 1976), 
and Holweger-M\"uller (1974) models.
The former is optimised for chromospheric
studies while the latter has
traditionally been the model of choice for solar abundance analyses.

More recently, a new generation of time-dependent 
2D and 3D hydrodynamical model atmospheres
has become available and applied in stellar abundance analyses
(e.g. Nordlund \& Dravins 1990; Dravins \& Nordlund 1990;
Steffen et al. 1995; Atroshchenko \& Gadun 1994; Stein \& Nordlund 1998;
Asplund et al. 1999, 2000a,b, 2004a,b; Asplund 2000, 2004; 
Asplund \& Garc\'{\i}a P{\'e}rez 2001;
Steffen \& Holweger 2002; V\"ogler et al. 2004).
By simultaneously solving the standard hydrodynamical equations for conservation of
mass, momentum and energy together with the 3D radiative transfer equation, both
the radiative and convective energy transport are self-consistently computed.
No mixing length theory with its free parameters therefore enters the modelling.
The effects of line-blanketing are accounted for through the use of opacity binning
assuming LTE
(Nordlund 1982). State-of-the-art equation-of-state (e.g. Mihalas et al. 1988)
and opacities (e.g. Kurucz 1993; Cunto et al. 1993) are used.
One can easily imagine that for late-type stars such as the Sun where the
convection zone reaches up to and even beyond the optical surface, such 3D hydrodynamical
modelling is necessary to capture the time-dependence and resulting atmospheric
inhomogeneities of granulation.
Being more sophisticated does not, however, necessarily make such 3D models more
realistic. A great deal of effort lately has therefore been invested in testing
the predictions of these models against an arsenal of observational constraints,
in particular for the Sun.
So far the comparison has been very satisfactory, which gives confidence
in the realism of the 3D models and their predictions (e.g. Stein \& Nordlund 1998).
For example, the intrinsic
Doppler shifts inherent in the hydrodynamical modelling render the
classical concepts of micro- and macroturbulence obsolete, yet the theoretical
3D line profiles agree essentially perfectly with the observed profiles, including
their line shifts and asymmetries (e.g. Asplund et al. 2000a).

Besides the atmospheric modelling, it is also necessary to accurately model the
line formation and the radiative and collisional processes that couple the radiation field
with the gas. Most analyses are based on the assumption of LTE, which simplifies
matter immensely by assuming that the atomic populations are described fully by
the Boltzmann and Saha distributions for excitations and ionizations.
In reality, for conditions typical of stellar atmospheres, the collisional rates
do not dominate over the corresponding radiative rates, which is a pre-requisite
for LTE to hold. Often it is therefore important to account for departures from
LTE, which can be substantial in terms of derived abundances.
The numerical techniques and codes to compute the statistical equilibrium in 1D
have been available for decades, indeed even 3D non-LTE calculations are
now feasible (e.g. Asplund et al. 2003).
In spite of this, remarkably little work has been devoted to this
important aspect of abundance analyses even in 1D.
In fact, to date only a dozen or so elements have had
their non-LTE line formation in solar-type atmospheres
studied in some detail with contemporary atomic physics.
Perhaps this can blamed on a notion within the astronomy community that the
input physics is very uncertain.
This is, however, only partly true, since the last decade has seen
tremendous progress in this respect. Furthermore, it is possible to test the
sensitivity of the outcome to the various rates, which can constrain
the non-LTE effects. It is important to realize that LTE 
is more likely than not to be wrong with departures from LTE 
often being on the level of 0.1\,dex for the Sun.
Depending on the required accuracy in the derived abundances, LTE may or may not
therefore be an acceptable approximation.

\section{The Solar Photospheric Chemical Composition}

The solar abundance analysis which is described below has been carried out
using a 3D hydrodynamical model of the solar atmosphere (Asplund et al. 2000a).
Both atomic and molecular lines of very different temperature and pressure
sensitivities have been employed whenever possible in the hope that it should minimise
systematic errors. A special effort has been made to utilise
only the best available line data. In several incidences detailed non-LTE
calculations have been carried out or the results thereof been culled from
the literature. As a result, 
we are confident that the end-product is the most reliable
element abundance analysis of the Sun carried out so far.
Table \ref{t:sun} presents a compilation of the
most reliable solar and meteoritic abundances in our opinion.

\begin{table}[!ht]
\caption{Element abundances in the present-day solar photosphere and in meteorites
(C1 chondrites).
Indirect solar estimates are marked with [..]
\label{t:sun}}
\smallskip
\begin{tabular}{llcc|llcc}
 \tableline
 \noalign{\smallskip}
 & Elem.  &  Photosphere  &   Meteorites &
 & Elem.  &  Photosphere  &   Meteorites  \\
 \noalign{\smallskip}
 \tableline
 \noalign{\smallskip}
1  & H   & $12.00$            & $8.25 \pm 0.05$         & 44 & Ru  &  $1.84 \pm 0.07$   &  $1.77 \pm 0.08$  \\
2  & He  & $[10.93 \pm 0.01]$   & $1.29$                & 45 & Rh  &  $1.12 \pm 0.12$   &  $1.07 \pm 0.02$  \\
3  & Li  &  $1.05 \pm 0.10$   &  $3.25 \pm 0.06$        & 46 & Pd  &  $1.69 \pm 0.04$   &  $1.67 \pm 0.02$  \\
4  & Be  &  $1.38 \pm 0.09$   &  $1.38 \pm 0.08$        & 47 & Ag  &  $0.94 \pm 0.24$   &  $1.20 \pm 0.06$  \\
5  & B   &  $2.70 \pm 0.20$     &  $2.75 \pm 0.04$      & 48 & Cd  &  $1.77 \pm 0.11$   &  $1.71 \pm 0.03$  \\
6  & C   &  $8.39 \pm 0.05$   &  $7.40 \pm 0.06$        & 49 & In  &  $1.60 \pm 0.20$   &  $0.80 \pm 0.03$  \\
7  & N   &  $7.78 \pm 0.06$   &  $6.25 \pm 0.07$        & 50 & Sn  &  $2.00 \pm 0.30$   &  $2.08 \pm 0.04$  \\
8  & O   &  $8.66 \pm 0.05$   &  $8.39 \pm 0.02$        & 51 & Sb  &  $1.00 \pm 0.30$   &  $1.03 \pm 0.07$  \\
9  & F   &  $4.56 \pm 0.30$   &  $4.43 \pm 0.06$        & 52 & Te  &  $$                &  $2.19 \pm 0.04$  \\
10 & Ne  &  $[7.84 \pm 0.06]$   &  -$1.06$                & 53 & I   &  $$                &  $1.51 \pm 0.12$  \\
11 & Na  &  $6.17 \pm 0.04$   &  $6.27 \pm 0.03$        & 54 & Xe  &  $[2.27 \pm 0.02]$   &  -$1.97$  \\
12 & Mg  &  $7.53 \pm 0.09$   &  $7.53 \pm 0.03$        & 55 & Cs  &  $$                &  $1.07 \pm 0.03$  \\
13 & Al  &  $6.37 \pm 0.06$   &  $6.43 \pm 0.02$        & 56 & Ba  &  $2.17 \pm 0.07$   &  $2.16 \pm 0.03$  \\
14 & Si  &  $7.51 \pm 0.04$   &  $7.51 \pm 0.02$        & 57 & La  &  $1.13 \pm 0.05$   &  $1.15 \pm 0.06$  \\
15 & P   &  $5.36 \pm 0.04$   &  $5.40 \pm 0.04$        & 58 & Ce  &  $1.58 \pm 0.09$   &  $1.58 \pm 0.02$  \\
16 & S   &  $7.14 \pm 0.05$   &  $7.16 \pm 0.04$        & 59 & Pr  &  $0.71 \pm 0.08$   &  $0.75 \pm 0.03$  \\
17 & Cl  &  $5.50 \pm 0.30$   &  $5.23 \pm 0.06$        & 60 & Nd  &  $1.45 \pm 0.05$   &  $1.43 \pm 0.03$  \\
18 & Ar  &  $[6.18 \pm 0.08]$   &  -$0.45$                & 62 & Sm  &  $1.01 \pm 0.06$   &  $0.92 \pm 0.04$  \\
19 & K   &  $5.08 \pm 0.07$   &  $5.06 \pm 0.05$        & 63 & Eu  &  $0.52 \pm 0.06$   &  $0.49 \pm 0.04$  \\
20 & Ca  &  $6.31 \pm 0.04$   &  $6.29 \pm 0.03$        & 64 & Gd  &  $1.12 \pm 0.04$   &  $1.03 \pm 0.02$  \\
21 & Sc  &  $3.05 \pm 0.08$   &  $3.04 \pm 0.04$        & 65 & Tb  &  $0.28 \pm 0.30$   &  $0.28 \pm 0.03$  \\
22 & Ti  &  $4.90 \pm 0.06$   &  $4.89 \pm 0.03$        & 66 & Dy  &  $1.14 \pm 0.08$   &  $1.10 \pm 0.04$  \\
23 & V   &  $4.00 \pm 0.02$   &  $3.97 \pm 0.03$        & 67 & Ho  &  $0.51 \pm 0.10$   &  $0.46 \pm 0.02$  \\
24 & Cr  &  $5.64 \pm 0.10$   &  $5.63 \pm 0.05$        & 68 & Er  &  $0.93 \pm 0.06$   &  $0.92 \pm 0.03$  \\
25 & Mn  &  $5.39 \pm 0.03$   &  $5.47 \pm 0.03$        & 69 & Tm  &  $0.00 \pm 0.15$   &  $0.08 \pm 0.06$  \\
26 & Fe  &  $7.45 \pm 0.05$   &  $7.45 \pm 0.03$        & 70 & Yb  &  $1.08 \pm 0.15$   &  $0.91 \pm 0.03$  \\
27 & Co  &  $4.92 \pm 0.08$   &  $4.86 \pm 0.03$        & 71 & Lu  &  $0.06 \pm 0.10$   &  $0.06 \pm 0.06$  \\
28 & Ni  &  $6.23 \pm 0.04$   &  $6.19 \pm 0.03$        & 72 & Hf  &  $0.88 \pm 0.08$   &  $0.74 \pm 0.04$  \\
29 & Cu  &  $4.21 \pm 0.04$   &  $4.23 \pm 0.06$        & 73 & Ta  &  $$                &  -$0.17 \pm 0.03$ \\
30 & Zn  &  $4.60 \pm 0.03$   &  $4.61 \pm 0.04$        & 74 & W   &  $1.11 \pm 0.15$   &  $0.62 \pm 0.03$  \\
31 & Ga  &  $2.88 \pm 0.10$   &  $3.07 \pm 0.06$        & 75 & Re  &  $$                &  $0.23 \pm 0.04$  \\
32 & Ge  &  $3.58 \pm 0.05$   &  $3.59 \pm 0.05$        & 76 & Os  &  $1.45 \pm 0.10$   &  $1.34 \pm 0.03$  \\
33 & As  &  $$                &  $2.29 \pm 0.05$        & 77 & Ir  &  $1.38 \pm 0.05$   &  $1.32 \pm 0.03$  \\
34 & Se  &  $$                &  $3.33 \pm 0.04$        & 78 & Pt  &                    &  $1.64 \pm 0.03$  \\
35 & Br  &  $$                &  $2.56 \pm 0.09$        & 79 & Au  &  $1.01 \pm 0.15$   &  $0.80 \pm 0.06$  \\
36 & Kr  &  $[3.28 \pm 0.08]$   &  -$2.27$                & 80 & Hg  &  $$                &  $1.13 \pm 0.18$  \\
37 & Rb  &  $2.60 \pm 0.15$   &  $2.33 \pm 0.06$        & 81 & Tl  &  $0.90 \pm 0.20$   &  $0.78 \pm 0.04$  \\
38 & Sr  &  $2.92 \pm 0.05$   &  $2.88 \pm 0.04$        & 82 & Pb  &  $2.00 \pm 0.06$   &  $2.02 \pm 0.04$  \\
39 & Y   &  $2.21 \pm 0.02$   &  $2.17 \pm 0.04$  & 83 & Bi  &  $$                &  $0.65 \pm 0.03$  \\
40 & Zr  &  $2.59 \pm 0.04$   &  $2.57 \pm 0.02$  & 90 & Th  &  $$                &  $0.06 \pm 0.04$  \\
41 & Nb  &  $1.42 \pm 0.06$   &  $1.39 \pm 0.03$  & 92 & U   &  $<$-$0.47       $   &  -$0.52 \pm 0.04$  \\
42 & Mo  &  $1.92 \pm 0.05$   &  $1.96 \pm 0.04$  & \\
 \noalign{\smallskip}
 \tableline
\end{tabular}
\end{table}

\subsection{Lithium, Beryllium and Boron}

Lithium is depleted in the solar convection zone by about a factor of 160 and as
a consequence even the Li\,{\sc i} resonance line at 670.8\,nm is very weak
(equivalent width $W_\lambda \simeq 0.18$\,pm=1.8\,m\AA ); 
the subordinate line at 610.4\,nm is
hopelessly puny for the purpose of deriving a solar Li abundance.
In addition, the 670.8\,nm line is blended with CN and Fe\,{\sc i} lines,
which makes the determination of the solar Li abundance rather uncertain.
Furthermore, Li\,{\sc i} is a minority species, which is highly sensitive to both the
temperature and the photo-ionizing UV radiation field (Kiselman 1997; Uitenbroek 1998).
Our recommended solar Li abundance stems from the 1D LTE analysis by M\"uller et al. (1975).
To this value
we add the effects of atmospheric inhomogeneities and 3D non-LTE line formation
(Asplund et al. 1999; Asplund et al. 2003; Barklem et al. 2003) and correct
for a revised transition probability to arrive at
$\log \epsilon_{\rm Li} = 1.05\pm0.10$.

Until recently Be was thought to be depleted in the solar photosphere by about a factor
of two (Chiemlewski et al. 1975; Anders \& Grevesse 1989). Balachandran \& Bell (1998), however, revisited
the issue by attempting to estimate any missing UV opacity and found a Be abundance
from the Be\,{\sc ii} 313.1\,nm resonance doublet
in agreement with the meteoritic value. There is a long-standing debate
whether or not
there is a need for additional continuous and/or line opacity in the UV spectral region
over those already included in the model calculations
in order to properly model the solar flux distribution.
The novel feature with the analysis of Balachandran \& Bell was their estimation of
the amount of missing UV opacity by requiring that the OH A-X electronic lines around 313\,nm
should yield the same solar oxygen abundance as the OH vibration lines in the IR.
In order to achieve this, they estimated that about 60\% extra opacity was required, which
they attributed largely to photo-ionization of Fe\,{\sc i} (see also Bell et al. 2001).
A follow-up study using a 3D solar model atmosphere instead of the {\sc marcs}
(Gustafsson et al. 1975) and Holweger-M\"uller (1974) 1D models employed by Balachandran \& Bell
confirmed that Be is undepleted in the solar convection zone:
$\log \epsilon_{\rm Be} = 1.38\pm0.09$ (Asplund 2004).
It should be emphasized, however, that this result hinges completely on the correctness
of the utilised method of estimating the missing opacity. In particular,
departures from LTE for the OH A-X electronic lines could skew the results.
No sufficiently detailed non-LTE calculation exists for these lines to our knowledge,
which obviously is important in order to settle the issue.
Likewise, improved determinations of the Fe\,{\sc i} bound-free cross-sections would
be very valuable.

The problem with possible missing UV opacity may also affect the determination of the solar B abundance,
as the only available reliable indicator is the B\,{\sc i} 249.67\,nm resonance line.
Cunha \& Smith (1999) presented a re-analysis of the B\,{\sc i} line, including a careful
study of possible missing opacity sources such as photo-ionization of Mg\,{\sc i}.
Whilst the uncertainty in their result was relatively large, they
found no evidence for a solar B depletion, which of course would be highly surprising if
indeed Be is undepleted.
We have estimated the 3D LTE effect (i.e. 3D LTE - 1D LTE) on the resonance line to be
0.04\,dex in comparison with a solar {\sc marcs} or Kurucz model and $-0.04$\,dex for
the Holweger-M\"uller model. When also including departures from LTE as predicted by
Kiselman \& Carlsson (1996), we obtain
$\log \epsilon_{\rm B} = 2.70\pm0.20$.
The large uncertainty is dominated by the problem in analysing this very crowded spectral region.

\subsection{Carbon}

\begin{table}[!t]
\caption{The derived solar C, N and O abundances as implied by a variety
of different atomic and molecular indicators using a 3D hydrodynamical
model of the solar atmosphere (Asplund et al. 2004a,b; Asplund et al., in preparation;
Scott et al., in preparation).
In addition, the corresponding results
for two 1D model atmospheres, a theoretical {\sc marcs} (Asplund et al. 1997)
and the semi-empirical model of Holweger-M\"uller (1974), are given for
comparison purposes. Note the excellent agreement between atomic and molecular
diagnostics with the 3D model and the large spread when using 1D model
atmospheres. The quoted uncertainty is only the line-to-line scatter;
no error is given for [C\,{\sc i}] as the result is based only on the 872.7\,nm line.
The O abundance used for the CO calculations was 8.66 in the case of 3D 
and 8.85 for the two 1D models
\label{t:CNO}
}
\smallskip
\begin{center}
\begin{tabular}{lccc}
 \tableline
 \noalign{\smallskip}
 lines  & \multicolumn{3}{c}{${\rm log} \epsilon_{\rm C, N, O}$} \\
 \noalign{\smallskip}
 \cline{2-4}
 \noalign{\smallskip}
          &   3D & HM & {\sc marcs} \\
 \noalign{\smallskip}
 \tableline
 \noalign{\smallskip}
$$[C\,{\sc i}] &  $8.39 $~~~~~~~~~       & $8.45 $~~~~~~~~~         & $8.40 $~~~~~~~~~ \\
C\,{\sc i}   & $8.36 \pm 0.03$ & $8.39 \pm 0.03$ & $8.35 \pm 0.03$ \\
CH $\Delta v = 1$   & $8.38 \pm 0.04$ & $8.53 \pm 0.04$ & $8.42 \pm 0.04$ \\
C$_2$ Swan   & $8.44 \pm 0.03$ & $8.53 \pm 0.03$ & $8.46 \pm 0.03$ \\
CH A-X       & $8.45 \pm 0.04$ & $8.59 \pm 0.04$ & $8.44 \pm 0.04$ \\
CO $\Delta v = 1$  & $8.41 \pm 0.02$ & $8.62 \pm 0.02$ & $8.57 \pm 0.02$ \\
CO $\Delta v = 2$  & $8.38 \pm 0.02$ & $8.70 \pm 0.03$ & $8.59 \pm 0.03$ \\
 \noalign{\smallskip}
 \tableline
 \noalign{\smallskip}
N\,{\sc i}   & $7.85 \pm 0.08$ & $7.97 \pm 0.08$ & $7.94 \pm 0.08$ \\
NH $\Delta v = 1$   & $7.73 \pm 0.05$ & $7.95 \pm 0.05$ & $7.82 \pm 0.05$ \\
 \noalign{\smallskip}
 \tableline
 \noalign{\smallskip}
[O\,{\sc i}] & $8.68 \pm 0.01$ & $8.76 \pm 0.02$ & $8.72 \pm 0.01$ \\
O\,{\sc i}   & $8.64 \pm 0.02$ & $8.64 \pm 0.08$ & $8.72 \pm 0.03$ \\
OH $\Delta v = 0$   & $8.65 \pm 0.02$ & $8.82 \pm 0.01$ & $8.83 \pm 0.03$ \\
OH $\Delta v = 1$   & $8.61 \pm 0.03$ & $8.87 \pm 0.03$ & $8.74 \pm 0.03$ \\
 \noalign{\smallskip}
 \tableline
\end{tabular}
\end{center}
\end{table}

The solar carbon abundance
can be determined from a wide range of indicators, both atomic and molecular transitions.
We have recently performed a detailed study of the best abundance diagnostics
in our opinion: the forbidden [C\,{\sc i}] 872.7\,nm line, high excitation C\,{\sc i} lines,
CH vibration-rotation IR lines, CH A-X electronic lines, C$_2$ Swan electronic lines and
CO vibration lines 
(Allende Prieto et al. 2002; Asplund et al. 2004b; Scott et al., in preparation).
The line formation was computed based on a 3D hydrodynamical model of the solar atmosphere
(Asplund et al. 2000a) and departures from LTE were accounted for in the case of
the C\,{\sc i} lines. In sharp contrast with analyses using 1D model atmospheres,
very gratifying agreement is found between all abundance indicators when employing the 3D model,
as clear from Table \ref{t:CNO}.
Given the extreme temperature sensitivity of molecule formation, it is crucial to
include the effects of atmospheric inhomogeneities for the molecular lines in order
not to estimate too high abundances, which is the case with 1D models.
Our new solar C abundance of $\log \epsilon_{\rm C} = 8.39\pm0.05$ is 0.17\,dex
lower than the recommended value in the widely used compilation of Anders \& Grevesse (1989).
The excellent agreement between transitions of very different formation depths and
temperature and pressure sensitivities is a very strong argument in favour of this low
abundance as well as for the realism of the 3D model.
In particular we note with satisfaction that consistent results are now finally provided also
by the CO lines, which have previously caused a great deal of trouble when analysed
within the framework of 1D model atmospheres (Grevesse et al. 1995; Ayres 2002 and references 
therein).

A 3D analysis of weak $^{12}$C$^{16}$O and $^{13}$C$^{16}$O lines yields a photospheric
$^{12}$C/$^{13}$C ratio of $95\pm5$ (Scott et al., in preparation),
which is in excellent agreement with telluric
measurements (Rosman \& Taylor 1998).

\subsection{Nitrogen}

We have recently re-determined the solar N abundance on the basis of the same
3D hydrodynamical model atmosphere as previously used for the
corresponding C and O analyses (Asplund et al, in preparation).
In the case of N there are fewer abundance indicators
compared with its two neighboring elements.
We have utilised N\,{\sc i} and NH vibration-rotation lines while relegating
the CN red band to a supporting role.
The N\,{\sc i} list consists of 21 weak ($W_\lambda = 0.1-1$\,pm),
high-excitation ($\chi_{\rm exc} = 10.3-12.4$\,eV) lines, which implies a 3D LTE
abundance of $\log \epsilon_{\rm N} = 7.88\pm0.08$.
No 3D non-LTE study for N is as yet available but the application of the
computed 1D non-LTE abundance corrections by Rentzsch-Holm (1996) lowers this abundance to
$\log \epsilon_{\rm N} = 7.85\pm0.08$. It should be noted, however, that
Rentzsch-Holm (1996) employs efficient inelastic H collisions, which help thermalise the
level populations. Given that the available laboratory and quantum mechanical
calculations tend to suggest that the classical recipe
used by her significantly over-estimates the H collision cross-sections
(e.g. Belyaev et al. 1999; Barklem et al. 2003),
the N\,{\sc i} based abundance presented here may require a
small downward revision.

We have also used a sample of 22 weak NH vibration-rotation lines in the IR
(located around $3 \mu$m), which have previously been advocated as primary abundance
indicators (Grevesse et al. 1990). 
As with all molecular lines, the NH lines are very temperature
sensitive and as consequence there is a large difference between the results
of 1D and 3D analyses. The NH lines indicate an abundance of
$\log \epsilon_{\rm N} = 7.73\pm0.05$ with our 3D solar model, which is significantly
lower than with the standard Holweger-M\"uller 1D model.

\subsection{Oxygen}

Allende Prieto et al. (2001) argued for a significantly lower solar O abundance
than the then widely accepted value, largely due to the presence of
a previously unaccounted for blend in the forbidden [O\,{\sc i}] 630.0\,nm line.
We have subsequently performed a much more detailed study, which also included
high-excitation permitted O\,{\sc i} lines and molecular transitions of OH, both
vibration-rotation and pure rotation lines (Asplund et al. 2004a).
As the O\,{\sc i} lines are known to
be susceptible to non-LTE effects, full 3D non-LTE line formation calculations
were performed for these lines.
It has long been clear that the OH lines imply higher abundance by about 0.2\,dex
than the O\,{\sc i} lines when including departures from LTE,
regardless of which 1D model atmosphere has been used.
Finally this problem has been resolved with the application of 3D hydrodynamical
models. The agreement between the various atomic and molecular lines is
excellent, in spite of their very different atmospheric sensitivities.
The almost perfect match between predicted and observed line profiles and
center-to-limb variations provides further evidence that the
3D model is highly realistic.
Furthermore, Melendez (2004) finds a similarly low O abundance from the
extremely weak first overtone OH vibration lines, which have previously not been used.

A 3D analysis of isotopic CO lines yields a photospheric
$^{16}$O/$^{18}$O ratio of $530\pm40$ (Scott et al., in preparation),
which is in good agreement with the telluric
data (Rosman \& Taylor 1998).

\subsection{Neon and Argon}

The photospheric abundances of Ne and Ar can not be determined 
directly due to the lack of suitable spectral lines.
Their recommended values in Table \ref{t:sun} have been
estimated from measured abundance ratios in the solar corona
and solar energetic particles in conjunction with the photospheric
abundance for the reference element, which in this case is O. 
The Ne and Ar abundances are therefore directly affected by the revised
solar O abundance. This explains most of the difference compared
with for example Anders \& Grevesse (1989) with a secondary effect
coming from refined coronal abundance ratio determinations
(Reames 1999). 

\subsection{Sodium to Calcium}

In connection with the present conference, we have performed
a preliminary 3D LTE abundance analysis for the remaining elements up
to calcium. We have been heavily influenced by the
study of Lambert \& Luck (1978) who performed a similar 
analysis using the Holweger-M\"uller (1974) 
semi-empirical 1D model atmosphere. In particular our choice
of lines has been largely moulded by their recommendations,
with the added constraint that we have so far restricted the
comparison to only weak lines. Line profile
fitting has been used in all cases
to estimate the abundance instead of equivalent
widths. We have ransacked the recent literature
for the most up-to-date transition probabilities. 
As appropriate, we take into account departures from LTE, 
although computed for 1D models since full 3D non-LTE 
calculations for these elements are well beyond the scope of
the present study. While this is inconsistent
in principle it is expected to give at least a good first estimate
of the 3D non-LTE effects (Asplund et al. 2004a). 
Space permits only a very brief summary here with the full details
being presented elsewhere (Asplund et al., in preparation).
The preliminary results are listed in Table \ref{t:sun}.
The new 3D-based abundance are in all cases slightly lower
than those recommended by Anders \& Grevesse (1989) and
Grevesse \& Sauval (1998), the difference being typically
$0.05-0.10$\,dex. The impact of the 3D model atmosphere is
thus smaller here than for C, N and O, mainly since these
abundances are based on atomic transitions rather
than very temperature sensitive molecular lines. 

The Na abundance is based on six weak Na\,{\sc i} lines. 
The computed non-LTE abundance corrections are small ($<0.05$\,dex) but slightly
lower the mean abundance to $\log \epsilon_{\rm Na} = 6.17\pm0.04$,
which is 0.1\,dex lower than the meteoritic value. As the
$gf$-values should be reliable, an explanation for this minor
discrepancy has not yet been identified.
The transition probabilities for Mg\,{\sc i} are notoriously uncertain,
forcing the Mg abundance to be estimated from four Mg\,{\sc ii} lines:
$\log \epsilon_{\rm Mg} = 7.53\pm0.09$. Non-LTE
effects should be negligible for Mg\,{\sc ii}. The Mg\,{\sc i} value
is very similar but with unacceptably large uncertainties. 
Many Al\,{\sc i} lines are detected in the solar spectrum but
the majority give quite discrepant results, mainly due 
to cancellation and configuration interaction effects in the
calculations of the $gf$-values. We consider only four Al\,{\sc i}
lines to be of sufficiently high quality for this purpose, which
suggests $\log \epsilon_{\rm Al} = 6.37\pm0.06$. 
Based on the published departure coefficients by 
Baum\"uller \& Gehren (1996),
possible non-LTE effects should be minor for these
relatively high-excitation lines.
The Si abundance is that of Asplund (2000). 

Our recommended P abundance stems from an analysis of five P\,{\sc i}
lines in the near-IR with equivalent widths $W_\lambda < 2.5$\,pm. 
With the $gf$-values of Berzinsh et al. (1997) we find 
$\log \epsilon_{\rm P} = 5.36\pm0.04$.
From the line list used by Lambert \& Luck (1978), we removed
several S\,{\sc i} lines with profiles that indicate blends. 
The pruned list of ten lines indicate $\log \epsilon_{\rm S} = 7.14\pm0.05$.
The new S abundance is now in excellent agreement with the meteoritic value.
Chlorine, like fluorine, 
has not been possible to analyse with our 3D model atmosphere
of quiet Sun as the relevant HCl lines are only present in sunspot spectra.
The highly uncertain Cl abundance dates back to Hall \& Noyes (1972).
As we have restricted the analysis to weak and intermediate 
strong lines, we have not yet used the very 
strong K\,{\sc i} 769.8\,nm resonance line and instead base our
K abundance on five weak K\,{\sc i} transitions
at 404.4, 580.1, 1176.9, 1243.2 and 1252.2\,nm.
Together with the non-LTE abundance corrections of 
Ivanova \& Shimanskii (2000), these lines imply 
$\log \epsilon_{\rm K} = 5.08\pm0.07$.
Finally, for Ca a nice selection of both permitted 
and forbidden Ca\,{\sc i} transitions as well as
Ca\,{\sc ii} lines are available in the solar spectrum.
The agreement between the two ionization stages is excellent,
and we simply take the average of the two:
$\log \epsilon_{\rm Ca} = 6.31\pm0.04$.

\subsection{Other Elements}

Of the remaining elements, only Fe has been the topic of a
3D abundance analysis (Asplund et al. 2000b), whose value
we adopt here: $\log \epsilon_{\rm Fe} = 7.45\pm0.05$.
Almost identical abundances are suggested by Fe\,{\sc i}
and Fe\,{\sc ii} lines. We note that this value has
not been corrected for departures from LTE.
Shchukina \& Trujillo Bueno (2001) found that the Fe\,{\sc i}
based abundance should be adjusted upwards by about 0.05\,dex
based on 1.5D (i.e. only including vertical radiative transfer
and therefore treating each atmospheric column as a separate 1D model 
atmosphere) non-LTE calculations using one snapshot from
the 3D solar model used here. Their calculations did not
include any inelastic H collisions, which would tend to decrease
the non-LTE effects towards restoring the LTE results. 
The jury is still out whether or not
these collisions are efficient. Korn et al. (2003) present
convincing arguments that they are in the case of Fe,
and we therefore tentatively recommend the 3D LTE value. 
Clearly, detailed quantum mechanical calculations for
Fe+H collisions are very important but unfortunately the
task is extremely challenging. 

While no 3D studies have been carried out for the remaining elements,
the recommended values in Table \ref{t:sun} nevertheless differ
in several incidences compared with previous compilations. 
This is largely the result of the very important work done
by a small group of dedicated atomic physicists in measuring and
calculating improved transition probabilities. 
Much recent attention has been devoted to the heavy elements
formed by neutron capture, like La (Lawler et al. 2001a),
Nd (Den Hartog et al. 2003), Eu (Lawler et al. 2001c), 
Tb (Lawler et al. 2001b), Ho (Lawler et al. 2004),
Pt (Den Hartog et al. 2004) and Pb (Bi\'{e}mont et al. 2000).
Some beautiful examples of this are shown in Sneden et al. (these proceedings).
Some other elements whose abundances have been revised due
to improved atomic data and/or line selection in recent years are
Ti, Ni (Reddy et al. 2003), Ge (Bi\'{e}mont et al. 1999)
and Sr (Barklem \& O'Mara 2000).

Finally, we note that we have not here attempted to
homogenise the rather diverse collection of photospheric abundances
presented in Table \ref{t:sun}. 
They have not all been derived using the same model atmosphere, 
departures from LTE have only occasionally been accounted for and
the methods for estimating the attached uncertainties differ
between authors. This should be borne in mind by the prospective users 
of this data.

\section{Comparison with the Meteoritic Evidence}

\begin{figure}[!t]
\plotone{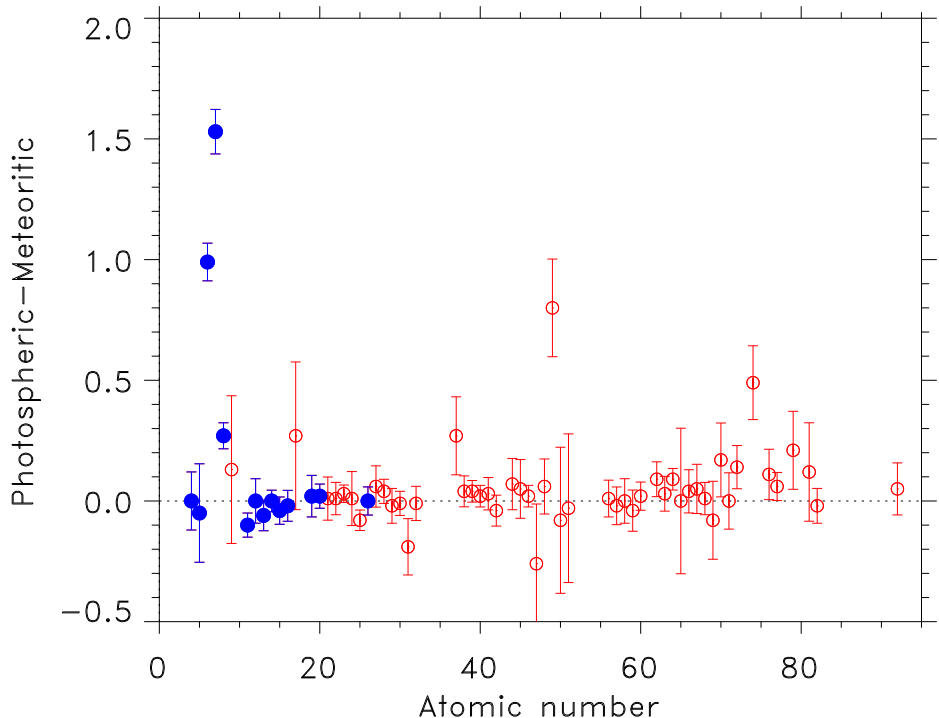}
\caption{Comparison of photospheric and meteoritic abundances (as
measured in C1 chondrites). The elements which have been 
analysed using a 3D hydrodynamical solar model atmosphere are
shown as filled circles while the 1D-based results with open circles.
H, Li and the noble gases all fall outside the figure due to
significant depletion either in the Sun or in the meteorites, 
which also affects C, N and O
\label{f:sunmet}}
\end{figure}

In Table \ref{t:sun} we also present a compilation of meteoritic 
abundances as measured in C1 chondrites. The data is largely taken
from Lodders (2003) but placed on a slightly different absolute
abundance scale. 
It is important to realize that since H, like other volatile
elements, is highly depleted in meteorites, another reference
element must be found. The honour is normally
given to Si. As a consequence, in order to anchor the meteoritic 
and photospheric absolute abundance scales to each other a
knowledge of the photospheric Si abundance is required. Since our 
recommended value is 0.03\,dex lower than that advocated by
Lodders (2003), we correspondingly 
adjust all meteoritic abundances by this amount.

Fig. \ref{f:sunmet} shows the difference between the 
photospheric and meteoritic abundances. 
In general the agreement is very good. 
Some notable differences can be easily explained by depletion in the
Sun (Li) or in the meteorites (H, C, N, O and the noble gases). 
In addition there are a small number of elements which show 
discordant results: Ga, Rb, Ag, In and W, with Cl and Au only barely
agreeing within the large uncertainties. In the case of
In and W one may suspect unidentified blends being the cause
of the discrepancies given that the photospheric abundances
are significantly higher than the meteoritic values. 
For the other problematic elements erroneous $gf$-values
or departures from LTE may be to blame. 
As mentioned above, the new Na abundance
does not overlap with the meteoritic value. 
Although the difference is only 0.1\,dex, this deserves
further investigation.  
The mean difference for the non-depleted elements is
$0.01\pm0.06$\,dex, when ignoring the above-mentioned seven elements.

\section{Trouble in Paradise?}

The new solar abundances described here have some important ramifications.

The first one is evident from the abundances by mass of H, He and metals, 
$X$, $Y$ and $Z$, which
now become $X = 0.7392$, $Y = 0.2486$ and $Z = 0.0122$ with $Z/X = 0.0165$,
i.e. much lower metallicity and $Z/X$ than previously recommended 
(e.g. $Z/X = 0.0275$ from Anders \& Grevesse 1989). These are
the present-day photospheric values with the He abundance taken from
Basu \& Antia (2004).

The low solar C and O abundances are now in much better agreement
with those measured in the local interstellar medium 
(e.g. Andr\'e et al. 2003) 
as well as for nearby B stars (e.g. Sofia \& Meyer 2001).
Previously the high solar abundances apparently already
in place 4.5\,Gyr ago were inconsistent with the low values found
in the solar neighborhood today in the framework of Galactic
chemical evolution. Somewhat surprisingly, the problem appears
to have been in the solar determination rather than in the
analyses of nebulae and hot stars. 

While the new abundances may have solved some long-standing problems,
it also introduces at least one new, namely for helioseismology.
Over the years, standard solar interior models which account for
element settling have yielded an impressive agreement with
the observed sound speed in the Sun as measured by the solar oscillations.
Since C, N, O and Ne are important opacity contributors, the 
revised solar abundances for these elements significantly alter
the interior structure. 
Indeed the predicted sound speed in these new solar models 
are in much worse agreement with helioseismology 
(Bahcall et al. 2004; Basu \& Antia 2004).
According to these authors, the only possible culprits are the
new photospheric abundances or the opacities, having discarded
the possibility of underestimated element settling in the solar models.
The necessary opacity increase near the bottom of the convection zone
is significant, at least 10\%. Some of this may in fact already have been
found (Seaton \& Badnell 2004), but more than half still remains. 
Whether the fault lies entirely with the opacity calculations
or whether our new abundances are in fact underestimated is an open question,
which urgently needs to be resolved. 
Experience constantly reminds us that one should never claim
anything as impossible, but we consider it unlikely that our 
derived abundances are as far from the real values as suggested
by these helioseismology studies.
It would be a remarkable conspiracy of factors if this was the case,
considering that the atomic and molecular-based abundances now finally
agree, the lack of any significant trends in derived abundances with
for example line strength or excitation potential and the almost
perfect match between observed and predicted line profiles, including
their asymmetries. 
The employed lines in our work probe very different parts of the solar
atmospheric with greatly different sensitivities to the atmospheric
conditions, which implies that a great deal of fine-tuning would be
necessary to simultaneously bring the C, N and O abundances up 
by some 0.2\,dex, if at all possible. 
As mentioned above, the new abundances also removes the special nature
the Sun otherwise would have in comparison with its neighborhood.

\section{Concluding Remarks}

Although progress has been made in refining our knowledge of the
solar chemical composition, much work still remains.
An obvious way forward is to extend the 3D analysis presented here
to the remaining elements in the periodic table. 
Furthermore, the lack of detailed non-LTE line formation calculations is
a major cause of concern, not only for the solar abundance determinations
but also for much of the stellar work.
Equally important,
however, is continued investment in improving the necessary atomic
and molecular data.  
Although the impact may not be as profound as 
the recent surprisingly large revisions of the solar metal content,
efforts along these lines would nevertheless be worthwhile.

\acknowledgments{It is a great pleasure to pay tribute to David Lambert for all of his
many important contributions to stellar spectroscopy and his unique ability to
bridge the gap to the nuclear and atomic physics communities.
Among his many achievements we would like to here high-light his several seminal papers
on the solar chemical composition, which to a great extent have shown the way for
very accurate abundance determinations. We would also like to thank our various
collaborators related to this work, including
Carlos Allende Prieto, Paul Barklem, Ronny Blomme, Mats Carlsson,
Dan Kiselman, David Lambert, \AA ke Nordlund, Pat Scott, Bob Stein
and Regner Trampedach.
MA and NG gratefully acknowledge financial support from
the SOC/LOC. 

\end{document}